\documentclass[12pt]{article}
\usepackage{amsmath,amssymb,bm}
\usepackage{cite}  
\usepackage{hyperref}
\hypersetup{colorlinks=true}
\title{Gravitational lensing by Reissner-Nordstr\"{o}m black holes with topological defects }
\date{}
\author{\bf Kimet Jusufi\footnote{E-mail: kimet.jusufi@unite.edu.mk}}

\makeatletter
 
  \@addtoreset{equation}{section}
 \makeatother

\begin{document}
\maketitle

\centerline{\it Department of Physics, State University of Tetovo, Ilinden Street nn, 1200, Macedonia}

\vskip 0.5 truecm

\abstract{Using a new geometrical approach introduced by Gibbons and Werner we study the deflection angle by Reissner-Nordstr\"{o}m black holes in the background spacetimes with a global monopole and a cosmic string. By calculating the corresponding optical Gaussian curvature and applying the Gauss-Bonnet theorem to the optical metric we find the leading terms of the deflection angle in the weak limit approximation. We find that the deflection angle increases due to the presence of topological defects.}

\section{Introduction}

One of the most important predictions of Einstein's General Theory of Relativity is the deflection of light rays from a distant source passing close to a massive body. This phenomena is crucial and played a key role as a first experimental verification of general
relativity.

Besides the standard methods used for calculating the deflection angle, recently, a new spin forward was put by Gibbons and Werner \cite{gibbons} by introducing the optical geometry and applying the Gauss-Bonnet theorem to the optical geometry. Using this approach, one can show the importannce of topology in gravitational lensing, stated in other words the focusing of the light rays can be viewed globaly as topological effect of the spacetime.  Along this line of research a number of static and spherically symmetric spacetime metrics have been studied \cite{gibbons,gibbons1}. Recently, this method was extended to stationary metrics for the Kerr black hole solution \cite{werner}.

Topological defects, such as domain walls, cosmic strings and monopoles, may have been produced by the phase transition in the early universe due to the breakdown of local or global gauge symmetries \cite{kible}. Cosmic strings are one dimensional object, characterized by the tension $G\mu$, where $G$ is the Newton's gravitational constant and $\mu$ is the mass per unit length of the string. It is interesting to notice that athought globally the spacetime around  a cosmic string is conical, localy the spacetime is just Minkowski spacetime. One way to think about this spacetime is to consider a Minkowski spacetime with a wedge removed given by the deficit angle $\delta=8\pi G \mu $. Note that, in the case of a global monopole the spacetime is not locally flat. However, globally the spacetime of the surface $\theta=\pi/2$, is conical, with deficit angle $\Delta=8\pi^{2}G\eta^{2}$ \cite{birola}.

In this context, is important to ask whether there is a chance that may lead to a possible
experimental detection of topological defects. There are however two classes of effects that may lead to indirect experimental evidence: the quantum effects and gravitational effects. Among these effects, a cosmic string can act as a gravitational lens \cite{gott}, it can induce a finite electrostatic self-force on an electric charged particle \cite{linet}, shifts in the energy levels of a hydrogen atom \cite{bezerra}, they were also suggested as an explanation of the anisotropy of the cosmic microwave background radiation, and many others. 

The aim of this paper is to use Gauss-Bonnet theorem and calculate the deflection angle around the spacetime of the Reissner-Nordstr\"{o}m black holes in the background with a global monopole and a cosmic string. It is widely believed that charged black holes don't exist in nature for a long time, since any initial charge will rather quickly be neutralized. However, as wa discussed here \cite{punsley}, charged black holes may be produced during the gravitational collapse of massive magnetized rotating stars surrounded by a co-rotating magnetosphere of equal and opposite charge. The magnetosphere preserves the black hole from a neutralization due to a selective accretion of charge from the environment and the black hole can be quite stable in a typical astrophysical environment of low density. Gravitational lensing from charged black holes in the strong field scenario has been discussed by \cite{eiroa}, in the case of the weak limit the deflection angle was also studied by \cite{sereno}, using Fermat's principle.

This paper is organized as follows. In Section 2, we briefly review the Gauss-Bonnet theorem and then calculate the corresponding Gaussian optical curvature for Reissner-Nordstr\"{o}m black hole. In Section 3, we calculate the leading terms of the deflection angle by Reissner-Nordstr\"{o}m black hole with topological defects. In Section 4, we comment on our results. We use metric signature $ (-,+,+,+) $ and $c=G=1$. The Latin indices are used to donate spatial coordinates and Greek indices for spacetime coordinates.

\section{Reissner-Nordstr\"{o}m optical metric with topological defects}
\label{sec:1}
We can start by writing the simplest Lagrangian density, which describes a global monopole given by 
\begin{equation}
\mathcal{L}=\frac{1}{2}g^{\mu \nu}\partial_{\mu}\phi^{a}\partial_{\nu}\phi_{a}-\frac{1}{4}\lambda (\phi^{a}\phi_{a}-\eta^{2})^{2},
\end{equation}
where $\lambda$ is the self-interaction term  and $\eta$ is scale of gauge-symmetry breaking $\eta\sim 10^{16} $ GeV, $\phi^{a}$ is a triplet of scalar fields which transform under the group $O(3)$, whose symmetry is spontaneously broken to $U(1)$ given by
\begin{equation}
\phi^{a}=\eta h(r)\frac{x^{a}}{r}
\end{equation}
with $x^{a}x^{a}=r^{2}$. The most general static metric with spherical symmetry can be written as
\begin{equation}
\mathrm{d}s^{2}=-A(r)\mathrm{d}t^{2}+A^{-1}(r)\mathrm{d}r^{2}+r^{2}\left(\mathrm{d}\theta^{2}+\sin^{2}\theta \,\mathrm{d}\varphi^{2} \right).\label{metric1}
\end{equation}

Solving the Einstein field equations in spacetime with a global monopole \cite{mello}, leads to the following expression for $A(r)$
\begin{equation}
A=1-8\pi \eta^{2}-\frac{2M}{r}+\frac{Q^{2}}{r^{2}},
\end{equation}
where $M$ is the black hole mass and $Q$ black hole charge. By introducing the following coordinate transfromations $r\to(1-8\pi\eta^{2})^{-1/2}r$,  $t\to(1-8\pi\eta^{2})^{1/2}t$ and $M\to(1-8 \pi \eta^{2})^{-3/2}M$, $Q\to (1-8\pi \eta^{2})Q$ \cite{vilenkin}, and also  by introducing a cosmic string in the metric \eqref{metric1}, which can be done by using $\varphi\to(1-4\mu)\varphi$, we end up with the following Reissner-Nordstr\"{o}m  black hole metric in spherical coordinates given by
\begin{equation}
\mathrm{d}s^{2}=-\left(1-\frac{2M}{r}+\frac{Q^{2}}{r^{2}}\right)\mathrm{d}t^{2}+\left(1-\frac{2M}{r}+\frac{Q^{2}}{r^{2}}\right)^{-1}\mathrm{d}r^{2}+a^{2}r^{2}\left(\mathrm{d}\theta^{2}+p^{2}\sin^{2}\theta \mathrm{d}\varphi^{2}\right),
\label{rn}
\end{equation}
where the terms $a^{2}=1-8\pi\eta^{2}$ and $p^{2}=(1-4\mu)^{2}$, encodes the presence of a global monopole and a cosmic string, respectively. The metric \eqref{rn}, describes a static and noninteracting infinitely long cosmic string and a global monopole placed close to each other in Reissner-Nordstr\"{o}m black hole spacetime.  An infinitely long cosmic string, is aligned through $\theta=0$ and $\theta=\pi/2$, parallel to $z$-axes, passing through the Reissner-Nordstr\"{o}m black hole spacetime. Without loss of generality, we can consider the equatorial plane $\theta=\pi/2$, in this way solving for null geodesics with $\mathrm{d}s^{2}=0$, the optical line element becomes 
\begin{eqnarray}
\mathrm{d}t^{2}=\frac{\mathrm{d}r^{2}}{\left(1-\frac{2M}{r}+\frac{Q^{2}}{r^{2}}\right)^{2}}+\frac{(1-8\pi \eta^{2})(1-4\mu)^{2}r^{2}\mathrm{d}\varphi^{2}}{1-\frac{2M}{r}+\frac{Q^{2}}{r^{2}}}.
\label{metric2}
\end{eqnarray}

For convenience, we can write the optical metric $\tilde{g}_{ab}$ in terms of the new coordinate $r^{\star}$ as
\begin{equation}
\mathrm{d}t^{2}= \tilde{g}_{ab}\,\mathrm{d}x^{a}\mathrm{d}x^{b}=\mathrm{d}{r^{\star}}^{2}+f^{2}(r^{\star})\mathrm{d}\varphi^{2},
\label{optical}
\end{equation}
where the function  $f(r^{\star})$ is given by 
\begin{equation}
f(r^{\star})=\frac{\sqrt{1-8\pi\eta^{2}}(1-4\mu)\,r}{\sqrt{1-\frac{2M}{r}+\frac{Q^{2}}{r^{2}}}}.\label{f}
\end{equation}

The corresponding optical metric from \eqref{optical}, therefore reads
\begin{equation}
\tilde{g}_{ab}=\begin{bmatrix}
    1      & 0 \\
    0       & f^{2}(r^{\star})  \\
\end{bmatrix},
\end{equation}
with the determinatnt $\det \tilde{g}_{ab}=f^{2}(r^{\star})$. Using the optical metric \eqref{optical}, it is not difficult to find that the only nonvanishing Christoffel symboles are $\Gamma^{r}_{\varphi\varphi}=-f(r^{\star})f'(r^{\star})$ and $\Gamma^{\varphi}_{r\varphi}=f'(r^{\star})/f(r^{\star})$. One can now proceed to calculate the corresponding Gaussian optical curvature $K$ as an intrinsic measure of curvature. In the case of 2-dimensional surface, this is not difficult since the only nonvanishing component of the Riemann tensor for the optical curvature is given by \cite{gibbons}
\begin{equation}
R_{r \varphi r \varphi}=K\left(\tilde{g}_{r \varphi}\tilde{g}_{\varphi r}-\tilde{g}_{rr}\tilde{g}_{\varphi \varphi}\right)=-K\det \tilde{g}_{r\varphi}
\end{equation}
where $R_{r\varphi r \varphi}=\tilde{g}_{r r}\,R^{r}_{\varphi r \varphi}$. It follows that the Gaussian optical curvature is given by
\begin{equation}
K=-\frac{R_{r\varphi r\varphi}}{\det \tilde{g}_{r \varphi}}=-\frac{1}{f(r^{\star})}\frac{\mathrm{d}^{2}f(r^{\star})}{\mathrm{d}{r^{\star}}^{2}}.
\end{equation}

Using the last equation, one can show that the intrinsic Gaussian optical curvature $K$, can be expressed in terms of $r$ as \cite{bao}
\begin{eqnarray}\label{Gcurvature}
K&=&-\frac{1}{f(r^{\star})}\frac{\mathrm{d}^{2}f(r^{\star})}{\mathrm{d}{r^{\star}}^{2}}=-\frac{1}{f(r^{\star})}\left[\frac{\mathrm{d}r}{\mathrm{d}r^{\star}}\frac{\mathrm{d}}{\mathrm{d}r}\left(\frac{\mathrm{d}r}{\mathrm{d}r^{\star}}\right)\frac{\mathrm{d}f}{\mathrm{d}r}+\left(\frac{\mathrm{d}r}{\mathrm{d}r^{\star}}\right)^{2}\frac{\mathrm{d}^{2}f}{\mathrm{d}r^{2}}\right].
\end{eqnarray}

Finally, we can now compute the corresponding Gaussian optical curvature for the Reissner-Nordstr\"{o}m black hole in presence of topological defects by substituting the equation \eqref{f} into \eqref{Gcurvature}, and find the following expression
\begin{equation}
K=-\frac{2M}{r^{3}}\left(1-\frac{3M}{2r}\right)+\frac{3Q^{2}}{r^{4}}\left(1+\frac{2Q^{2}}{3r^{2}}\right)-\frac{6MQ^{2}}{r^{5}}.
\label{curvature}
\end{equation}

Now it's interesting to see that Gaussian optical curvature $K$ seems to be independent of topological defects, but as we will see in the next section, there is a contribution due to the topology of spacetime which is globally conical. The other interesting point about the last equation which may come as a surprise is that, $K$ is negative, which implies that locally light rays should diverge. As was pointed out by \cite{gibbons}, the negative Gaussian curvature of the optical metric is a rather general feature of black hole metrics, and the interesting thing here is that, only globally, by considering the topology of spacetime, light rays can converge. As we will see, this can be done by using the Gauss-Bonnet theorem.

\section{Deflection angle by Reissner-Nordstr\"{o}m black hole with topological defects}

Having said that, let us now write the Gauss-Bonnet theorem, which relates the intrinsic geometry of the spacetime with its topology, for the region $D_ {R} $ in $M$, with boundary $\partial D_{R}=\gamma_{\tilde{g}}\cup C_ {R} $ which states that (see, e.g.,\cite{gibbons})
\begin{equation}
\int\limits_{D_{R}}K\,\mathrm{d}S+\oint\limits_{\partial D_{R}}\kappa\,\mathrm{d}t+\sum_{i}\epsilon_{i}=2\pi\chi(D_{R}),
\end{equation}
where $K$ is the Gaussian curvature, $\kappa$ the geodesic curvature, given by
$ \kappa=\tilde{g}\,(\nabla_{\dot{\gamma}}\dot{\gamma}, \ddot{\gamma})$, such that $\tilde{g}(\dot{\gamma}, \dot{\gamma})=1$, with the unit acceleration vector $\ddot{\gamma}$ and $\epsilon_{i}$ the corresponding exterior angle at the $i$\,th vertex. As $R\to \infty$, both jump angles become $\pi/2$, in other words $\theta_{O}+\theta_{S}\to \pi$. Since $D_ {R} $ is non-singular, than the Euler characteristic is $\chi(D_{R})=1$, finally we are left with
\begin{equation}
\iint\limits_{D_{R}}K\,\mathrm{d}S+\oint\limits_{\partial D_{R}}\kappa\,\mathrm{d}t=2\pi\chi(D_{R})-(\theta_{O}+\theta_{S})=\pi.
\label{gaussbonnet}
\end{equation}

Since $\gamma_{\tilde{g}}$ is geodesic, clearly than $\kappa(\gamma_{\tilde{g}})=0$, the only interesting part to be calculated remains $\kappa(C_{R})=|\nabla_{\dot{C}_{R}}\dot{C}_{R}|$ as $R\to \infty$. It should be noted that, unlike the case of optical Schwarzschild metric which is asymptotically Euclidean, i.e. $\kappa(C_{R})\mathrm{d}t/\mathrm{d}\varphi=1$, this is not valilid if topological defects are introduced. To see this, let's calculate at very large $R$, given by  $C_{R}:= r(\varphi)=R=const.$, the radial component of the geodesic curvature
\begin{equation}
\left(\nabla_{\dot{C}_{R}}\dot{C}_{R}\right)^{r}=\dot{C}_{R}^{\varphi}\,\partial_{\varphi}\dot{C}_{R}^{r}+\Gamma^{r}_{\varphi \varphi}\left(\dot{C}_{R}^{\varphi}\right)^{2},
\end{equation}
we can use the fact that $\tilde{g}_{\varphi \varphi}\,\dot{C}_{R}^{\varphi}\dot{C}_{R}^{\varphi}=1$, which implies $\dot{C}_{R}^{\varphi}=1/f^{2}(r^{\star})$, and recall that $\Gamma^{r}_{\varphi\varphi}=-f(r^{\star})f'(r^{\star})$, yielding
\begin{equation}
\left(\nabla_{\dot{C}_{R}^{r}}\dot{C}_{R}^{r}\right)^{r}=-\frac{f'(r^{\star})}{f(r^{\star})}\to-\frac{1}{R}.
\end{equation}

We can now use the last result and compute the geodesic curvature 
\begin{equation}
\kappa(C_{R})=|\nabla_{\dot{C}_{R}}\dot{C}_{R}|=\left(\tilde{g}_{rr}\dot{C}^{r}_{R}\dot{C}^{r}_{R}\right)^\frac{1}{2}\to\frac{1}{R}.
\end{equation}

At very large $r(\varphi)=R=const.$, it follows that the geodesic curvature is  indipendent of topological defects, $\kappa(C_{R})\to R^{-1}$, however   from the optical metric \eqref{optical}, it's not difficult to see that $\mathrm{d}t=(1-4\mu)\sqrt{1-8\pi\eta^{2}}\,R \,\mathrm{d}\,\varphi$, and hence 
\begin{equation}
\kappa(C_{R})\mathrm{d}t=\frac{1}{R}(1-4\mu)\sqrt{1-8\pi\eta^{2}}\,R \,\mathrm{d}\,\varphi.
\end{equation}

Unlike the case of black hole configuration without topological defects, here the optical metric is not asymptotically Euclidian, since the last result leads to $\kappa(C_{R})\mathrm{d}t/\mathrm{d}\varphi=(1-4\mu)\sqrt{1-8\pi\eta^{2}}\neq 1$, which is satisfied only when $\mu=\eta=0$. Now, we could go back to \eqref{gaussbonnet}, and substitute this result, yielding
\begin{eqnarray}\label{def1}
\iint\limits_{D_{R}}K\,\mathrm{d}S&+&\oint\limits_{C_{R}}\kappa\,\mathrm{d}t\overset{{R\to \infty}}{=}\iint\limits_{S_{\infty}}K\,\mathrm{d}S+(1-4\mu)\sqrt{1-8\pi\eta^{2}} \int\limits_{0}^{\pi+\hat{\alpha}}\mathrm{d}\varphi.
\end{eqnarray}

In the weak deflection limit  we may assume that the light ray is given by $r(t)=b/\sin\varphi $ at zeroth order, using \eqref{curvature} and \eqref{def1} it follows that the deflection angle is given by 
\begin{equation}
\hat{\alpha}=\frac{\pi-\pi(1-4\mu)\sqrt{1-8\pi\eta^{2}}}{(1-4\mu)\sqrt{1-8\pi\eta^{2}}}-\frac{1}{(1-4\mu)\sqrt{1-8\pi\eta^{2}}}\int\limits_{0}^{\pi}\int\limits_{\frac{b}{\sin \varphi}}^{\infty}K\,\sqrt{\det \bar{g}}\,\mathrm{d}r\,\mathrm{d}\varphi .
\label{angle}
\end{equation}
where $\mathrm{d}S=\sqrt{\det \tilde{g}}\,\mathrm{d}r\,\mathrm{d}\varphi$, is the optical surface area and  $\sqrt{\det\tilde{g}}=(1-4\mu)\sqrt{1-8\pi\eta^{2}} \,r$. The first term can be approximated as 
\begin{equation}
\frac{\pi-\pi(1-4\mu)\sqrt{1-8\pi\eta^{2}}}{(1-4\mu)\sqrt{1-8\pi\eta^{2}}}\simeq 4\mu \pi+4\pi^{2}\eta^{2}.
\end{equation}

Substituting the leading terms of the Gaussian curvature \eqref{curvature} into the last equation we find 
\begin{eqnarray}\nonumber
\int\limits_{0}^{\pi}\int\limits_{\frac{b}{\sin \varphi}}^{\infty}K\mathrm{d}S &\approx &\int\limits_{0}^{\pi}\int\limits_{\frac{b}{\sin \varphi}}^{\infty}\left(-\frac{2M}{r^{3}}+\frac{3Q^{2}}{r^{4}}\right)\sqrt{\det \tilde{g}}\mathrm{d}r\mathrm{d}\varphi\\\nonumber
&\approx &\sqrt{1-8\pi\eta^{2}}}{(1-4\mu)\left(-\frac{4M}{b}+\frac{3\pi Q^{2}}{4b^{2}}\right)
\label{angle}
\end{eqnarray}
where the impact parameter $b>>2M$. Putting all together, we end up with the expression for the deflection angle given by
\begin{equation}
\hat{\alpha}\simeq 4\mu \pi+4\pi^{2}\eta^{2}+\frac{4M}{b}-\frac{3\pi Q^{2}}{4b^{2}}.\label{deflection}
\end{equation}

We have therefore found that, the total deflection angle by Reissner-Nordstr\"{o}m black hole in background with topological defetcs, actually increses due to the presence of the first and second term. The first term gives the deflection angle due to the presence of a cosmic string, the second term gives the deflection angle due to the presence of a global monpole, the third term is the well known result for Schwarzschild black hole, and finally, the last term is coming due to the charge nature of the black hole. The light signal is propagating from the source $S$ to an observer $O$, such that both $S$ and $O$ lie on the same surface $\theta=\pi/2$. If $S$, $O$, global monopole and a cosmic string are perfectly aligned, it follows
\begin{equation}
\hat{\alpha}\simeq\left( 4\mu \pi+4\pi^{2}\eta^{2}\right)l(d+l)^{-1}+\frac{4M}{b}-\frac{3\pi Q^{2}}{4b^{2}},
\end{equation} 
where $d$ and $l$ are the corresponding distances from the monopole/cosmic string to the observer $O$, and to the source $S$, respectively. It should be noted that the increased value of the deflecton angle is rather small, of the order of $10$ arcsec, since $G\mu \simeq 10^{-6}$ and $G\eta^{2}\simeq 10^{-5}$, which makes the possible detection difficult, however as pointed out by \cite{vilenkin}, this effect is in the observable range. It is widely believed that global monopoles should move with some velocity $v$, in this particular case, we assume that a whole system consisted by a cosmic string and a global monopole is moving with some relativistic speed $v$, parallel to $z$-axes, with respect to the obsorver $O$. If this system pierces the Reissner-Nordstr\"{o}m spacetime, than the deflection angle by using the same arguments as \cite{vilenkin}, can be written as 
\begin{equation}
\hat{\alpha}_{1}=\gamma^{-1}\left(1-\textbf{n}\textbf{v}\right)^{-1}\hat{\alpha}_{0}+\frac{4M}{b}-\frac{3\pi Q^{2}}{4b^{2}}
\end{equation}
here $\textbf{n}$ is the unit vector along the line of sight, $\gamma=(1-v^{2})^{-1/2}$ and $\hat{\alpha}_{0}=\left( 4\mu \pi+4\pi^{2}\eta^{2}\right)l(d+l)^{-1}$. Now let us consider and discuss some special cases. Setting $Q=0$, from \eqref{rn} we recover the Schwarzschild black hole with topological defects
\begin{equation}
\mathrm{d}s^{2}=-\left(1-\frac{2M}{r}\right)\mathrm{d}t^{2}+\left(1-\frac{2M}{r}\right)^{-1}\mathrm{d}r^{2}+a^{2}r^{2}\left(\mathrm{d}\theta^{2}+p^{2}\sin^{2}\theta \mathrm{d}\varphi^{2}\right).
\end{equation}

This metric has the same characteristics as metric \eqref{rn}, with the corresponding Gaussian optical curvature $K=-2M (1-3M/2r)/r^{3}$. It also follows directly from \eqref{deflection} the deflection angle $\hat{\alpha}\simeq 4\mu \pi+4\pi^{2}\eta^{2}+4M/b$. On the other hand if $Q=M=\mu=0$, we recover the Minkowski spacetime pirced by a global monopole given by
\begin{equation}
\mathrm{d}s^{2}=-\mathrm{d}t^{2}+\mathrm{d}r^{2}+(1-8\pi \eta^{2})r^{2}\left(\mathrm{d}\theta^{2}+\sin^{2}\theta \mathrm{d}\varphi^{2}\right).
\end{equation}

The characteristics of this metric are well known, it describes a spacetime with a deficit solid angle, the area of a sphere of radius $r$ is not $4\pi r^{2}$, but $4\pi(1-8\pi \eta^{2})r^{2}$. In the particular case of $\pi/2$, the surface has the geometry of a cone with a deficit angle $\Delta=8\pi^{2}\eta^{2}$. The Gaussian curvature is zero, $K=0$, so we end up with a deflection angle $\hat{\alpha}\simeq 4\pi^{2}\eta^{2}$ \cite{birola}. On the other hand if $M=Q=\eta=0$, we only have the contribution coming from a cosmic string, the metric in cylindrical coordinates reads \cite{aryal}
\begin{equation}
\mathrm{d}s^{2}=-\mathrm{d}t^{2}+\mathrm{d}z^{2}+\mathrm{d}r^{2}+(1-4\mu)^{2}r^{2}\mathrm{d}\varphi^{2}.
\end{equation}

This metric describes a static straight string lying along the $z$-axis. It's not difficult to see that for all surfaces $z=const.$, globally the topology is conical, with a deficit angle $\delta=8\pi \mu$. The Gaussian curvature is zero, $K=0$, with a deflection angle $\hat{\alpha}\simeq 4\mu \pi$ \cite{gibbons2}. And finally, in the spacetime without topological defects, $\eta=\mu=0$, the metric reduces to the Reissner-Nordstr\"{o}m black hole 
\begin{equation}
\mathrm{d}s^{2}=-\left(1-\frac{2M}{r}+\frac{Q^{2}}{r^{2}}\right)\mathrm{d}t^{2}+\left(1-\frac{2M}{r}+\frac{Q^{2}}{r^{2}}\right)^{-1}\mathrm{d}r^{2}+r^{2}\left(\mathrm{d}\theta^{2}+\sin^{2}\theta \mathrm{d}\varphi^{2}\right)
\end{equation}

The Gaussian curvature coincides with \eqref{curvature}, the deflection angle $\hat{\alpha}\simeq 4M/b-3\pi Q^{2}/4b^{2}$, is in agreement with \cite{sereno}. It will be interesting to see if this method can be generalised for the case of image multiplicity in a more general black hole configuration.

\section{Conclusion}
In this paper, we have investigated the deflection angle around Reissner-Nordstr\"{o}m black holes in the background spacetimes with topological defects. By adopting an optical metric and applying the Gauss-Bonnet theorem to the optical metric, we have found that the total deflection can be expressed as the sum of four corresponding terms $4\mu \pi, 4 \pi^{2}\eta^{2}$, $4M/b$ and $-3\pi Q^{2}/4b^{2}$, respectively. In the particular case $M=Q=0$, and $\eta=0$, the Reissner-Nordstr\"{o}m metric reduces to Minkowski spacetime pierced by an infinite static cosmic string with deflection angle $\hat{\alpha}=\delta/2=4\mu \pi$, and similary when $M=Q=\mu=0$, the metric reduces to Minkowski spacetime with a global monopole, with deflection angle $\hat{\alpha}=\Delta/2=4 \pi^{2}\eta^{2}$. Finally, setting $\eta=\mu=0$, we are left with the deflection angle around Reissner-Nordstr\"{o}m black hole $\hat{\alpha}=4M/b-3\pi Q^{2}/4b^{2}$. 

\section*{Acknowledgement}

The author would like to thank the anonymous reviewers for the very useful comments and suggestions which help us improve the quality of this paper.

\end{document}